\documentclass[journal]{IEEEtran}
\usepackage{cite}
\ifCLASSINFOpdf
   \usepackage[pdftex]{graphicx}
\else
   \usepackage[dvips]{graphicx}
\fi
\usepackage{amsmath}
\ifCLASSOPTIONcompsoc
  \usepackage[caption=false,font=normalsize,labelfont=sf,textfont=sf]{subfig}
\else
  \usepackage[caption=false,font=footnotesize]{subfig}
\fi
 \let\MYorigsubfloat\subfloat
 \renewcommand{\subfloat}[2][\relax]{\MYorigsubfloat[]{#2}}
\usepackage{bm}
\usepackage{multirow}
\allowdisplaybreaks

\usepackage{lineno}
\usepackage{booktabs,multirow,amsthm,amsmath,mathrsfs}
\usepackage{algorithm}
\usepackage{algorithmic} 
\usepackage{amssymb} 

\usepackage{subfig}
\usepackage{setspace}
\usepackage{changepage}

\usepackage{color}
\begin{document}

%\title{\LARGE{Modeling and Analysis Framework for Segment Routing with Computational Efficiency}}
%\title{\huge{Approximation and Online Algorithms for Multi-Source Multicast with Intra-Session Network Coding}}
\title{\huge{Extreme Flow Decomposition for Multi-Source Multicast with Intra-Session Network Coding}}


\author{Jianwei~Zhang%,~\IEEEmembership{Member,~IEEE}
	\thanks{This work was supported by the National Natural Science Foundation of China under Grant No. 61902346.
		\textit{(Corresponding author: Jianwei Zhang.)}
		%		, and the Zhejiang Provincial Natural Science Foundation of China under Grant No. XXXXXXX. 
		%		Corresponding author: Jianwei Zhang (janyway@outlook.com).
	}
	\thanks{Jianwei Zhang (janyway@outlook.com) is with the 
		School of Computer and Computing Science, 
		Zhejiang University City College, Hangzhou, China.}% <-this % stops a space
}

\markboth{MANUSCRIPT, 2020}%
{Shell \MakeLowercase{\textit{et al.}}: Bare Demo of IEEEtran.cls for IEEE Journals}



\maketitle

\begin{abstract}
%In this letter, we study the multicast streaming problem in a fully-connected hybrid wired/wireless overlay.
%We derive explicit formulas for the maximum streaming rate and the hybrid multiplicity using a flow-based method.
%The considered model supports both full-duplex and half-duplex modes,
%and incorporates equal service and differentiated service in a unified framework.
%The obtained results have theoretical and practical significance in wired/wireless content distribution.
%Online primal-dual algorithm
%
%Optimizing Network Throughput in Multi-Hop Wireless Networks with Segment Routing

Network coding (NC), when combined with multipath routing, enables a linear programming (LP) formulation for a multi-source multicast with intra-session network coding (MISNC) problem.
However, it is still hard to solve using conventional methods due to the enormous scale of variables or constraints.
In this paper,
we try to solve this problem in terms of throughput maximization from an algorithmic perspective.
We propose a novel formulation based on the extereme flow decomposition technique,
which facilitates the design and analysis of approximation and online algorithms.
For the offline scenario, we develop a fully polynomial time approximation scheme (FPTAS) which can find a $(1+\omega )$-approximation solution for any specified $\omega>0$.
For the online scenario, we develop an online primal-dual algorithm which proves $O(1)$-competitive and violates link capacities by a factor of $O(\log m)$, where $m$ is the link number.
The proposed algorithms share an elegant primal-dual form and thereby have inherent advantages of simplicity, efficiency, and scalability.
%We give worst case performance bounds for the proposed algorithms via rigorous proofs.
To better understand the advantages of the extereme flow decomposition approach, 
we devise delicate numerical examples on an extended butterfly network. 
We validate the effects of algorithmic parameters and make an interesting comparison between the proposed FPTAS and online algorithm.
The results show that the online algorithm has satisfactory performance while keeping the overall link utilization acceptable compared with the FPTAS.
\end{abstract}

\begin{IEEEkeywords}
%Multicast, Multi-commodity flow, Traffic engineering, Link utilization.
Network Coding, Multicast, Extreme Flow, Approximation Algorithm, Online Algorithm
\end{IEEEkeywords}

\IEEEpeerreviewmaketitle

\section{Introduction}

Among various potential benefits brought by \textit{Network Coding} (NC),
the most significant one is to maximize the multicast throughput in wired networks, or to improve the reliability and hence the unicast throughput in wireless neworks \cite{IEEE_Network_survey2019}.
NC and multipath routing are essential to formulate a multicast optimization problem with \textit{Intra-Session Network Coding} (ISNC) to a \textit{Linear Program} (LP) based on the classical \textit{Multi-Commodity Flow} (MCF) models \cite{cl}.
Even so, the LP is hard to solve using conventional methods due to the enormous scale of variables or constraints.
Naturally, this area of research divides into two major classes: \textit{Single-source multicast with ISNC} (SISNC) and \textit{Multi-source multicast with ISNC} (MISNC).

%Single NC multicast and \textit{Multicasts with Intra-Session Network Coding} (MISNC).

%
%
%Maximum concurrent flow problem for 
%involves a large scale of computation overheads.
%% maximum-multicommodity flow 
%As can be seen later, the primal-dual algorithms have elegant forms and have inherent advantages of simplicity, efficiency, and scalability.
%They are very simple to implement and run significantly faster than a general LP solver especially on medium and large sized problems. 
%

There are volumes of research outputs that optimize multicast performance or analyze the user behaviors in SISNC,
while few attention has been paid to the fundamental MISNC problem due to its computation complexity.
It is therefore significant to design efficient algorithms that can exactly or approximately solve the LP problem.

\begin{figure}[!h]
	\centering
	\includegraphics[angle=0, width=0.45\textwidth]{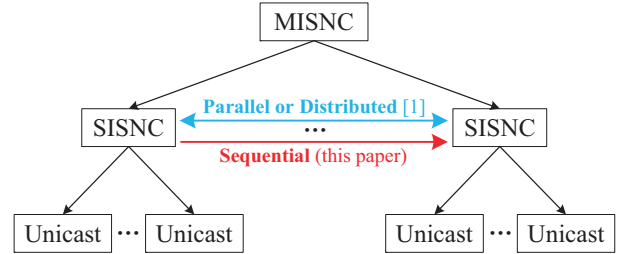}
	\caption{Comparison of approaches used in \cite{cl} and this paper.}
	\label{fg:method}
\end{figure}
In our prior work \cite{cl}, as shown in Fig.~\ref{fg:method}, we demonstrate that the original MISNC problem can be solved in a two-level decomposition framework.
At the upper level, a multi-source multicast is decomposed into a series of single-source multicast and at the lower level,
each single-source multicast is further decomposed into a series of unicast.
The solution is by nature an \textit{exact} algorithm features a design philosophy of \textit{trading time for space}.
In this paper, however, we investigate the MISNC problem from the perspective of how to efficiently obtain \textit{approximate} solutions, which features a design philosophy of \textit{trading accuracy for time and space}.
Moreover, the sequential processing structure enables an online implementation which has great realistic significance. 
Also, we focus on the upper level in this paper due to practical consideration and analytical tractability.
In summary, we aim to tackle the following challenges.
\begin{itemize}
	\item How to establish computationally tractable model to maximize the multicast throughput by leveraging the linear structural properties of ISNC?
	\item Further, how to design efficient algorithms to approximately solve the model with performance guarantees in both offline and online scenarios?
\end{itemize}

The problems are non-trivial due to the fact that
the MISNC problem cannot be formulated into path-based or tree-based models using conventional methods, and thus cannot be solved by adapting from existing algorithms designed for gereral MCF problems without NC.
More importantly, it is still unclear how to choose a set of multiple multicast trees that is optimal for the objective function \cite{ho}.
The reasons for this are two-fold:
first, the number of multicast trees in a network can be exponentially large;
second, the link sharing relationship between different trees caused by NC is intractable \cite{ho}.

Inspired by the facts that
\textit{a min-cost SISNC problem can be efficiently solved in a distributed or parallel manner \cite{cl,mit}}, and that 
\textit{a min-cost computation module is convenient to be integrated into a primal-dual scheme \cite{fptas2007,high_demands}},
we formulate the MISNC problem based on the \textit{extreme flow decomposition} technique \cite{high_demands} and develop primal-dual algorithms with guaranteed worst case performance for both offline and online settings.
Specifically, the contributions are as follows.
%The main contributions of this paper are the following:
\begin{itemize}
	\item We propose a novel optimization formulation based on the extereme flow decomposition for the MISNC problem, 
	which facilitates the design and analysis of approximation and online algorithms.
%	where segment number, SR-nodes number, intra-segment routing policy are all variables. 
%	It leads to a highly extensible framework to design and evaluate algorithms adaptive to various network topologies and traffic matrices.
	\item For the offline setting, we develop a \textit{fully polynomial time approximation scheme} (FPTAS) which can find a $(1+\omega )$-approximation solution for any specified $\omega>0$ in time that is a polynomial function of the network size.
%	\item For the online setting, we develop a \textit{practical} online algorithm and a modified \textit{theoretical} version with flow shifting. 
%	Both of them are designed in a primal-dual style and prove to be $O(1)$-competitive and violate link capacities by a factor of $O(\log m)$, where $m$ is the link number.
		\item For the online setting, we develop online algorithms with exact and approximate min-cost computation modules, respectively. 
	Both of them are designed in a primal-dual style and prove to be $O(1)$-competitive and violate link capacities by a factor of $O(\log m)$, where $m$ is the link number.
	\item We prove the performance bounds for the proposed algorithms and, to better understand the advantages of the extereme flow decomposition, we devise delicate numerical examples on a prototype network to illustrate the effects of the algorithmic parameters and make a direct comparison between the FPTAS and the online algorithm.
%	 , and validate the parameters of the proposed algorithms in both offline and online settings.
\end{itemize}

The rest of this paper is organized as follows. 
We review related work in Section II. 
We introduce the system model and assumptions in Section III.
We formulate the offline and online MISNC problems and develop approximation and online algorithms in Sections IV and V, respectively,
followed by the crucial min-cost computation modules in Section VI.
We then present numerical examples in Section VII. 
Finally, we conclude in Section VIII. 
All proofs are presented in the appendix. 
Main notation is summarized in Table~\ref{tb:notation}.

\section{Related Work}
\subsection{Intra-Session Network Coding (ISNC)}
ISNC is recently attracting public attention with the prolifiration of new generation of wireless networks.
Xu \textit{et al.} \cite{xu_edge} employ network coding and coalition game theory to jointly optimize the network throughput and content service satisfaction degree in mobile edge computing networks.
Wang \textit{et al.} \cite{iscc_d2d} propose a cooperative multicast for layered streaming in a heterogeneous D2D-enhanced cellular network.
Malathy \textit{et al.} \cite{IoTBackpressure} present a mixed routing scheme which combines network coding and backpressure routing in a large-scale IoT network.
In this scheme, network coding is used to minimize the network congestion, reduce the redundant packets, and enhance the throughput.
Zhu \textit{et al.} \cite{missForward} propose a Miss-and-Forward paradigm for ISNC, where a relay node is assigned to exploit the space diversity.
ISNC is also combined with the opportunistic routing to improve the reliability and throughput in multi-hop wireless networks \cite{IEEE_Network_survey2019}.

Regarding the wired networks,
existing works mainly focus on the SISNC problems partly because of the tractable scale of LP formulations for such case \cite{mit,li:ton,wu}.
These works can be roughly classified into two categaries.
First, all the receivers of a multicast are assumed to cooperate with each other to reach some global traffic engineering objectives, 
such as minimum cost, maximum throughput, minimum \textit{Maximum Link Utilization} (MLU).
Second, each multicast receiver is assumed to be selfish and greedily routes its flows, thus whether there exists an equilibrium
needs to be considered.

The considered MISNC problem falls into the first class and for this problem,
only heuristics \cite{gene}, approximation algorithms \cite{infocom:multicast} with no desirable performance guarantees, and suboptimal solutions \cite{comnet:multicast} are proposed.
In \cite{cl}, we formulated an LP for the MISNC problem in order to minimize the MLU.
We then presented a dual-dual decomposition approach \cite{alter} whereby the original problem is decomposed into a series of minimum-cost flow subproblems plus a centralized computation.
As such, the subproblems can be efficiently solved in a parallel or decentralized manner.
%Due to the effects of multipath and network coding, the obtained MLU can serve as a fundamental lower bound in traffic engineering.

%The reason for solving the problem as an FPTAS instead of a standard linear programming problem is that the FPTAS is very simple to implement and runs significantly faster than a general linear programming solver especially on medium and large sized problems. 
%software defined Mesh

%Unlike the above works, we use a novel ...

Essentially, the key components are all to solve a set of multicast optimization problems.
The decomposition in \cite{cl} occurs in spatial dimension, where multiple multicast requests can be processed simultaneously by different computation entities, or can of course be processed sequencially by the same computation entity.
In this paper, the decomposition occurs in temporal dimension, where multiple multicast requests must be processed sequencially,
although the order is not important. 
On the other hand, the decomposition method in \cite{cl} is by nature an exact algorithm which aims to trade time for space.
The centralized computation part involves a somewhat inefficient subgradient iteration module over a convex set.
The proposed algorithms in this paper are by nature approximation algorithms which sacrifice accuracy for computation efficiency.
We can use an appropriately chosen parameter to reach a trade-off in practice.
In both of the two decomposition methods, however, the updates of dual link weights at the upper level should be executed in a centralized manner.

\subsection{Primal-Dual Algorithms for Network Flow Problems}
The primal-dual scheme has been employed to solve a variety of MCF-based network flow problems \cite{model}, including both approximation algorithms and online algorithms.
According to the problem sturctures, different \textit{meta-structures}, i.e., the basic routing units, are taken into account, see Table~\ref{tb:comparison}.

%FPTAS
A long line of works have addressed the design of FPTAS for MCF under unicast scenarios \cite{fptas2007,fptas2008,fptas_sdn}. 
%
%P2P
For multicast, Sengupta \textit{et al.} \cite{p2p} develop a family of FPTASs that can maximize multicast streaming capacity and corresponding multicast trees in node capacitated P2P networks.
The algorithms involve smallest price tree construction in the innermost loops.
%the MCF problem with multiple multicast sessions is still a three-level structure.
%
% mesh, link/node protection
Kodialam \textit{et al.} \cite{mesh,mesh2} adapt the FPTAS to maximize the achievable rates in multi-hop wireless networks by jointly considering routing and scheduling problems.
The major difference is that the capacity constraint is specially designed for the wireless mesh nodes, and is flexible enough to handle various wireless interference conditions and channel models.
Bhatia \textit{et al.} \cite{protection} present FPTAS to maximize the throughput with fast restoration against link and node failures.
However, the generic path, used as the meta-structure, is no longer a simple path but consists of a primary path and a backup detour.
%
%SFC
Agarwal \textit{et al.} \cite{te:infocom} formulate the SDN controller optimization problem for traffic engineering with evolutinary deployment of SDN routing devices and develop an FPTAS for solving the problem.
In recent years, many works extend the shortest-path or min-cost tree based algorithms to design policy-aware routing algorithms that steer network flows through predefined service function chains under network function virtualizaiton environment in order to maximize the total flow demands \cite{steering,anu_unicast,anu_multicast}.
% SR
Bhatia \textit{et al.} \cite{sr} design FPTAS and online primal-dual algorithms to determine the optimal parameters for segment routing in both offline and online settings, and also consider the restoration from node or link failures \cite{sr_restoration}.

Unlike the above works, we treat each SISNC as an MCF, or equivalently, a feasible extreme flow in the flow polytope \cite{high_demands}.
In this way, the proposed FPTAS features a two-level (\textit{phase-iteration}) structure which outperforms the three-level (\textit{phase-iteration-step}) structures due to a substantial reduction in the number of loops as well as in running time.
The proposed online algorithm also benefits from this and thus achieve a disirable network throughput because multipath routing is inherently supported.

\section{System Model}

Let $G=(V,E)$ be a directed network, where $V$ is the node set and $E$ is the link set.
The number of nodes and links are denoted by $n$ and $m$, respectively.
There are a set of multicast requests $R$ to be transmitted through the network. 
Each request $r \in R$ is associated with one source $s_r \in V$ and a set $T_r \in V$ of receivers. 
We assume that all receivers in $T_r$ receive service at the same rate $d_r$ from $s_r$.

\begin{table}[!t]\small
	\caption{\label{tb:notation}Notations}
	\centering
	\begin{tabular}{p{1.6cm}p{6.4cm}}
		\toprule
		\textbf{Notation} & \textbf{Description} \\
		\midrule
		$G=(V, E)$ & A directed network where $V$ represents the node set and $E$ the link (edge) set. \\
		$n, m$ & Node number and link number of $G$, respectively. \\
		$r$ & Request $r$. \\
		$R, |R|$ & Request set $R$ and its cardinality. \\
		$s_r$, $T_r$ & Source node and target node set of request $r$. \\
		$d_r$ & Size of request $r$. \\
		$c_e$ & Capacity of link $e$. \\		
		${{\Pi }_{r}}$ & Flow polytope for request $r$. \\
		$V({\Pi _r})$ &  Set of extreme flows (extreme points) of ${{\Pi }_{r}}$. \\
		$f_e$ & Flow value that an extreme flow $f$ imposes on link $e$. \\
		%		$N_v$ & The set of links connected to node $v$. \\
		%		${P_r^k}$ & The SR-path through intermediate node $k$ due to request $r$. \\
		${x_{r}^{f}}$ & Flow amount of request $r$ routed as a scaled extreme flow $f$. \\
		${X_{r}^{f}}$ & Integer indicating whether a request $r$ has been accepted to route as a scaled extreme flow $f$. \\
		$x_e$ & Total flow value on link $e$. \\
		$p_e$ & Dual variable associated with each link $e$. \\
		$z_r$ & Dual variable associated with each request $r$. \\
		$\lambda$ & Maximum multiplier such that multicast rate $\lambda {{d}_{r}}$ can be supported for $r$. \\
		$\epsilon$ & Global tunable parameter of FPTAS. \\
		$\phi$ & Global tunable parameter of online algorithm. \\
		LP & Linear program. \\
		MCF & Multi-commodity flow. \\
		ISNC & Intra-session network coding. \\ 
		SISNC & Single-source multicast with ISNC. \\
		MISNC & Multi-source multicast with ISNC. \\
		FPTAS & Fully polynomial time approximation scheme. \\ 
		\bottomrule
	\end{tabular}
\end{table}

\subsection{Intra-Session Network Coding (ISNC)}
To focus on the major challenges, we make the following assumptions in this paper.
\begin{itemize}
	\item 
	Solving the problem includes two decoupled stages: 
	first, computing the routing subgraph; 
	second, determining the random coding scheme on the routing subgraph \cite{rnc,mit}.
	We concentrate on the former and do not go deep into the coding details.
	\item
	We assume that NC is only allowed within a multicast request (ISNC) and disallowed among different multicast requests because of the impracticability and computational intractability \cite{cl,li:ton}. 
\end{itemize}

Under the second assumption, the MISNC model can be viewed as a group of SISNC coupled by link capacity constriants.
The following theorem summarizes the most important property for an SISNC. 
 
\textbf{Theorem 1:}
In a directed network with network coding, a multicast rate is feasible \textit{if and only if} it is feasible to each receiver independently.
When using random linear coding schemes, there is a fully decentralized approach to achieve the min-cost multicast with strictly convex cost functions.
	
\textit{Proof:}
Refer to \cite{mit}.
\qed

\subsection{Extreme Flow Decomposition}
We adapt some definitions on the extreme flow decomposition in \cite{high_demands} to our problem as follows.

\textbf{Definition 1:} For each multicast request $r=({{s}_{r}},{{T}_{r}},{{d}_{r}})$, ${{\Pi }_{r}}$ denote the \textit{flow polytope} of unit flows $f$ (i.e., $|f|=1$) from $s_r$ to $T_r$ in $G$ that satisfy: ${{f}_{e}}\le \frac{{{c}_{e}}}{{{d}_{r}}},\forall e$. 
We say that request $r$ is feasible if ${{\Pi }_{r}}\ne \varnothing$.

\textbf{Definition 2:} The set of \textit{extreme flows (extreme points)} of ${{\Pi }_{r}}$ is denoted by $V({\Pi _r})$. 
Each extreme point represents a feasible flow which must saturates some links.
Any feasible flow can be represented by the linear combination of extreme flows. 

The comparison of various decomposition methods is given in Table~\ref{tb:comparison}.
When designing a primal-dual algorithm, the path or tree decomposition has a  three-level (\textit{phase-iteration-step}) structure,
while the extreme flow decomposition has a  two-level (\textit{phase-iteration}) structure.
\begin{table*}[!h]\small
	\begin{adjustwidth}{-1.0in}{-1.0in}
		\caption{\label{tb:comparison}Comparison of various decomposition methods}
		\centering
		\begin{tabular}{p{4.0cm}p{2.1cm}p{2.1cm}p{3.3cm}p{4.6cm}}
			\toprule
			\textbf{Strategy} & \textbf{Meta-structure} & \textbf{Routing} & \textbf{Computation} & \textbf{Feasible flow}\\
			\midrule
			\textbf{Path decomposition} & path & shortest path & unicast & sum of path flows\\
			\textbf{Tree decomposition} & tree & Steiner tree & multicast & sum of tree flows\\
			\textbf{Extreme flow decomposition} & extreme flow & min-cost MCF & unicast, multicast (ISNC) &  linear combination of extreme flows\\
			%			\textbf{MEP} & unnecessary & high & high & high & very easy\\
			\bottomrule
		\end{tabular}
	\end{adjustwidth}
\end{table*}

\section{Offline Network Throughput Maximization}
In the offline setting, we asume all the multicast requests are known in advance.
The objective is to simultaneously maximize the multicast throughput of all the requests.
We first present the offline formulation and then develop an approximation algorithm.
\subsection{Problem Formulation}
Similar to the well-kwown \textit{maximum concurrent flow problem} \cite{fptas2007}, 
the objective is to find the largest $\lambda$ such that there is an MCF which routes at least $\lambda {{d}_{r}}$ for all requests $r$.
Thus, the problem can be formulated as the following LP.
\begin{align}
\textbf{max}~&\lambda & (P_{\rm off}) \nonumber\\
%(P_{\rm off})~\textbf{max}~&\lambda \nonumber\\
\textbf{s.t.}~
\label{eq:poff:f} &\sum\limits_{f\in V({{\Pi }_{r}})}{{{y}_{f}}}\ge \lambda {{d}_{r}},&\forall r. \\
\label{eq:poff:c} &\sum\limits_{r}{\sum\limits_{f\in V({{\Pi }_{r}})}{{{f}_{e}}{{y}_{f}}}}\le {{c}_{e}},&\forall e. \\
&{{y}_{f}}\ge 0,&\forall r,\forall f\in V({{\Pi }_{r}}). \nonumber\\
&\lambda \in \mathbb{R}. \nonumber
\end{align}

%\begin{align}
%(P):~\textbf{max}~&\lambda \nonumber\\
%\textbf{s.t.}~
%\label{eq:p:f} \sum\limits_{k}{x_{r}^{k}}&=\lambda {{d}_{r}},\forall r. \\
%\label{eq:p:c} \sum\limits_{e\in N_v}{\frac{\sum\nolimits_{r}{\sum\nolimits_{k}{g_{r}^{k}(e)x_{r}^{k}}}}{{{c}_{e}}}}&\le 1,\forall v. \\
%x_r^k &\ge 0,\forall r,k \nonumber\\
%\lambda &\in \mathbb{R} \nonumber
%\end{align}

Similar to a typical MCF formulation, constraints (\ref{eq:poff:f}) and (\ref{eq:poff:c}) imply the flow conservation and capacity limitation, respectively.
By associating a price $p_e$ with each link $e$ and a weight $z_r$ with each request $r$,
we write the dual to the above LP as:
\begin{align}
\textbf{min}~&D(p):=\sum\limits_{e}{{{p}_{e}}{{c}_{e}}} & (D_{\rm off}) \nonumber\\
%(D_{\rm off})~\textbf{min}~&D(p):=\sum\limits_{e}{{{p}_{e}}{{c}_{e}}} \nonumber\\
\textbf{s.t.}~
\label{eq:poff:dual:p} &\sum\limits_{e}{{{p}_{e}}{{f}_{e}}}\ge {{z}_{r}},&\forall r,\forall f\in V({{\Pi }_{r}}). \\
\label{eq:poff:dual:z} &\sum\limits_{r}{{{z}_{r}}{{d}_{r}}}\ge 1. \\
&z_r \ge 0,&\forall r. \nonumber\\
&{p_e} \ge 0,&\forall e. \nonumber
\end{align}

Recall that a flow polytope may have exponentially many vertices.
Thus in the worst case, there could be at most exponential number of variables in $(P_{\rm off})$.
As far as we know, no tractable general-purpose LP solver can be directly applied to such problem even for a medium sized network. 

\subsection{Approximation Algorithm}

%We use a \textit{Fully Polynomial Time Approximation Scheme} (FPTAS) to solve the dynamic routing problem at the SDN-C.
%The reason for solving the problem as an FPTAS instead of a standard linear programming problem is that the FPTAS is very simple to implement and runs significantly faster than a general linear programming solver especially on medium and large sized problems. 
%An FPTAS provides the following performance guarantees: for any $\epsilon>0$, the solution has objective function value within ($1+\epsilon$)-factor of the optimal, and the running time is at most a polynomial function of the network size and $1/\epsilon$. 
%The FPTAS in our case is a primal-dual algorithm.

We design an FPTAS to solve the problem.
The FPTAS is a primal-dual algorithm which includes an outer loop of a primal-dual update and an inner loop of min-cost MCF path computation. 

The algorithm to solve the problem starts by assigning a precomputed price of $\frac{\delta }{{{c_e}}}$ to all links $e$.

The algorithm proceeds in phases.
In each phase, we route $d_r$ units of flow from node $s_r$ to $T_r$ for each multicast request $r$. 
A phase ends when all multicast requests are routed.

The flow of value $d_r$ of request $r$ is routed from $s_r$ to $T_r$ in multiple iterations as follows.
In each iteration, a min-cost MCF path with unit request size from $s_r$ to $T_r$ that minimizes the left-hand side of constraint (\ref{eq:poff:dual:p}) under current link prices is computed.

The path is computated in Section \ref{sec:kernel}.
The amount of flow sent along this path in an iteration is $d_r$.

%For a given SR-path t and prices p(), let Q(t, p) denote the Left-Hand-Side (LHS) of constraint (6), which we call the price of tree t. 
%A set of prices p() is a feasible solution for the dual program if and only if

After the flow of value $d_r$ is sent along the MCF path, the flow value and the link price at each link along the path $f^*$ are updated as follows:

1) Update the flow value $x_e$ as
\[{x_e} \leftarrow {x_e} + {f^*_e}{d_r}\]

2) Update the link prices $p_e$ as
\[{{p}_{e}}\leftarrow {{p}_{e}}\left( 1+\epsilon \frac{{f^*_e}{d_r} }{{{c}_{e}}} \right)\]

The update happens after each iteration associated with routing a complete request $d_r$. 
The algorithm terminates when the dual objective function value $D(p)$ becomes less than one.

When the algorithm terminates, dual feasibility constraints will be satisfied. 
However, link capacity constraints (\ref{eq:poff:c}) in the primal solution will be violated, since we were working with the original (not the residual) link capacities at each stage. 
To remedy this, we scale down the traffic at each link uniformly so that the link capacity constraints are satisfied.

Note that unlike the existing three-level FPTASs, FPTAS ISNC needs not to compute a bottleneck, or maximum allowable flow rate in the flow value update since in the \textit{Min-Cost SISNC Computation} module the link capacities have been scaled down by a factor $d_r$.

\textbf{Theorem 2:}
For any specified $\omega >0$, Algorithm \ref{alg_fptas} computes a $(1+\omega )$-approximation solution. 
If the algorithmic parameters are $\epsilon (\omega )=1-{{(1+\omega )}^{-\frac{1}{3}}}$ and $\delta (\omega )={{\left( \frac{1-\epsilon }{m} \right)}^{\frac{1}{\epsilon }}}$, 
the running time is $O\left( \frac{|R|\log |R|}{\epsilon }{{\log }_{1+\epsilon }}\frac{m}{1-\epsilon }{{{T_{{\rm{NC}}}}}} \right)$, 
where $T_{\rm{NC}}$ is the time required to compute a min-cost unit MCF for SISNC.

\textit{Proof:}
See Appendix.
\qed

\begin{algorithm}[!h]
	\caption{FPTAS for MISNC}\label{alg_fptas}
	\begin{algorithmic}
		\STATE \textbf{Input:} ${p_e} \leftarrow \frac{\delta }{{{c_e}}},\forall e$; 
		$x_e\leftarrow 0 ,\forall e$; 
		%		$F\leftarrow 0$;
		$\rho\leftarrow 0$
		%		\STATE Initialize ${f_{uv}(e)},\forall u,v \in {N_r} \cup \{ {s_r},{t_r}\} ,\forall r$ via the IGP routing policy using primal link weight system.
		\WHILE{$D(p) < 1$}	
		\FOR{$\forall r \in R$}
		\STATE Invoke \textbf{Algorithm \ref{alg_mcm}} to compute the min-cost unit MCF from $s_r$ to $T_r$ in $G$ under the link price system $p$. Denote $f^*$ as the min-cost unit MCF. 
		\FOR{$\forall e\in f^*$}
		\STATE ${x_e} \leftarrow {x_e} + {f^*_e}{d_r}$
		\STATE ${{p}_{e}}\leftarrow {{p}_{e}}\left( 1+\epsilon \frac{{f^*_e}{d_r} }{{{c}_{e}}} \right)$
		\ENDFOR
		\ENDFOR
		\STATE $\rho\leftarrow \rho+1$
		\STATE $D(p) \leftarrow \sum\limits_e {{p_e}{c_e}}$
		\ENDWHILE
		\STATE ${x_e} = \frac{{{x_e}}}{{{{\log }_{1 + \epsilon }}\frac{{1}}{\delta }}},\forall e$
%		\STATE ${x_e} = \frac{{{x_e}}}{{{{\log }_{1 + \epsilon }}\frac{{1 + \epsilon }}{\delta }}},\forall e$
		\STATE $\lambda  = \frac{{\rho - 1}}{{{{\log }_{1 + \epsilon }}\frac{1}{\delta }}}$
		\STATE \textbf{Output:} ${x_e}$; $\lambda$
	\end{algorithmic}
\end{algorithm}

%\textbf{Definition 3.}
%If the MCF generated by an approximation algorithm is $\alpha$-competitive and violates the link capacity constraints by $\beta$,  
%the online algorithm is said to be \textit{$\left\{ \alpha,\beta\right\}$-competitive}.
%Definition 1.1: An ?-approximation algorithm for an optimization problem is a polynomial-
%time algorithm that for all instances of the problem produces a solution whose value is within a
%factor of ? of the value of an optimal solution.
% ? the performance guarantee of the algorithm.
%In the literature, it is also often called the approximation ratio or approximation factor of the
%algorithm
% In this book we will follow the convention that ? > 1 for minimization problems,
%while ? < 1 for maximization problems. Thus, a
%1
%2
%-approximation algorithm for a maximization
%problem is a polynomial-time algorithm that always returns a solution whose value is at least
%half the optimal value.

%In this paper, it is also often called the approximation ratio or approximation factor of the
%algorithm
%
%Suppose that a primal-dual algorithm produces an $\alpha$-approximation solution and that

\section{Online Network Throughput Maximization}
%\subsection{Maximum (Achievable) Streaming Rate}
%In this section, we give the upper bound expression of the achievable streaming rate for $A$ in terms of the bandwidths of all nodes.
%In the proof, we develop a rate allocation and routing scheme to achieve the bound.
In the online setting, the multicast requests arrive one by one without the knowledge of future arrivals.
The objective is to accept as many multicast requests as possible.
We first formulate the online problem and then develop an online primal-dual algorithm.
\subsection{Problem Formulation}
Based on the  \textit{maximum multicommodity flow problem} \cite{fptas2007}, the online problem can be formulated as the following ILP.
\begin{align}
\textbf{max}~&\sum\limits_{r}{{{d}_{r}}\sum\limits_{f\in V({{\Pi }_{r}})}{X_{r}^{f}}}& (P_{\rm on}) \nonumber\\
%(P_{\rm on})~&\textbf{max}~\sum\limits_{r}{{{d}_{r}}\sum\limits_{f\in V({{\Pi }_{r}})}{X_{r}^{f}}}\nonumber\\
\textbf{s.t.}~
\label{eq:pon:f} &\sum\limits_{f \in V({\Pi _r})} {X_r^f}  \le 1,&\forall r. \\
\label{eq:pon:c} &\sum\limits_{r}{{{d}_{r}}\sum\limits_{f\in V({{\Pi }_{r}})}{{{f}_{e}}X_{r}^{f}}}\le {{c}_{e}},&\forall e. \\
&X_{r}^{f}\in \{ 0,1\},&\forall r,\forall f\in V({{\Pi }_{r}}). \nonumber
\end{align}

Similar to the offline formulation, constraints (\ref{eq:pon:f}) and (\ref{eq:pon:c}) imply the flow conservation and capacity limitation, respectively.
The integer $X_r^f$ indicates whether or not a request $r$ has been accepted to route as a scaled extreme flow $f$.
We then consider the LP relaxation of this problem where $X_{r}^{f}\in \{ 0,1\}$ is relaxed to $X_r^f \ge 0$. 
%\footnote{After the relaxation, the primal variables in problems $(P_{\rm on})$ and $(P_{\rm on})$ have the relation $x_r^f = {d_r}X_r^f$.}
Note that $X_r^f \le 1$ is already implied by constraint~(\ref{eq:pon:f}).
By associating a price $p_e$ with each link $e$ and a weight $z_r$ with each request $r$,
we can write the dual to the above LP as:
\begin{align}
\textbf{min}~&\sum\limits_r {{z_r}} {\rm{ + }}\sum\limits_e {{p_e}{c_e}}&(D_{\rm on}) \nonumber\\
%(D_{\rm on})~\textbf{min}~&\sum\limits_r {{z_r}} {\rm{ + }}\sum\limits_e {{p_e}{c_e}} \nonumber\\
\textbf{s.t.}~
\label{eq:pon:dual:c} &{z_r} \ge {d_r}\left( {1 - \sum\limits_e {{p_e}{f_e}} } \right),&\forall r,\forall f \in V({\Pi _r}). \\
&z_r \ge 0,&\forall r. \nonumber\\
&{p_e} \ge 0,&\forall e. \nonumber
\end{align}

\subsection{Online Algorithm}
%In the online setting, the routing requests arrive one by one without the knowledge of future arrivals.
We design an online primal-dual algorithm which includes an outer loop of a primal-dual update and an inner loop of min-cost MCF path computation. 

The algorithm to solve the problem starts by assigning a precomputed price of zero to all links $e$.

The algorithm proceeds in iterations and each iteration corresponds to a request.
Upon the arrival of a new request $r$, we try to route $d_r$ units of flow from node $s_r$ to $T_r$, for each multicast request $r$. 

In each iteration, a min-cost MCF path with unit request size from $s_r$ to $T_r$ that maximizes the right-hand side of constraint (\ref{eq:pon:dual:c}) under current link prices computed according to Section \ref{sec:kernel}.

If the min-cost value is larger than one, the request is rejected.
Otherwise, the request is accepted, and the entire flow $d_r$ of request $r$ is routed along the min-cost MCF path.

After the flow is sent, the flow value and the link price at each link along the path $f^*$ are updated as follows:

1) Update the flow value $X_r^{{f}}$ as
\[X_r^{{f^*}} \leftarrow 1,\]
which implies $X_r^f = 0,\forall f \ne {f^*}$ according to constraint (\ref{eq:pon:f}).

2) Update the link prices $p_e$ as
\[{p_e} \leftarrow {p_e}\left( {1 + \frac{{{f^*_e}{d_r}}}{{{c_e}}}} \right){\rm{ + }}\frac{\phi }{{m}}\frac{{{f^*_e}{d_r}}}{{{c_e}}},\phi >0\]

Parameter $\phi$ is designed to provide a tradeoff between the competitiveness of the online algorithm and the degree of violation of the capacity constraints. That is, a smaller $\phi$ leads to larger network throughput as well as a larger degree of violation on the link capacities \cite{anu_unicast}.
	
\textbf{Definition 3:}
If the MCF generated by an online algorithm is $\alpha$-competitive and violates the link capacity constraints by at most $\beta$,  
the online algorithm is said to be \textit{$\left\{ \alpha,\beta\right\}$-competitive}.
$\alpha$ and $\beta$ are also called \textit{competitive ratio} and \textit{violation ratio}, respectively.

\textbf{Theorem 3:}
Algorithm \ref{alg_online} is an all-or-nothing, non-preemptive, monotone, and $\left\{ {O\left( 1 \right),O\left( {\log m} \right)} \right\}$-competitive,  more specifically $\left\{ {{{1 + \phi }},\log \left( {\frac{{Bm}}{\phi } + 1} \right)} \right\}$-competitive, online algorithm.
	
	\textit{Proof:}
	See Appendix.
	\qed
	
\begin{algorithm}[!h]
	\caption{Online Algorithm for MISNC}\label{alg_online}
	\begin{algorithmic}
		\STATE \textbf{Input:} $p_e \leftarrow 0,\forall e$
		\WHILE{Upon the arrival of request $r$}	 
		\STATE Invoke \textbf{Algorithm \ref{alg_mcm}} to compute the min-cost unit MCF from $s_r$ to $T_r$ in $G$ under the link price system $p$. Denote $f^*$ as the min-cost unit MCF. 
		\IF{$L>1$}
		\STATE Reject $r$;
		\ELSE
		\STATE Accept $r$;	
		%		\STATE Let $k^* = \arg \mathop {\min }\limits_k \sum\limits_e {g_r^k(e)l(e)}$
		\STATE $X_r^{{f^*}} \leftarrow 1$ 
		%		\STATE ${z_r} \leftarrow {d_r}\left( {1 - \sum\limits_e {{p_e}{f^*_e}} } \right)$
		\STATE ${z_r} \leftarrow {d_r}\left( {1 - L } \right)$
		\FOR{$\forall e\in f^*$}
		%		\STATE \[\begin{aligned} 
		%		w_v \leftarrow w_v&\left( {1+ \sum\limits_{e \in N_v} {\frac{{g_r^k(e)d_r}}{{c_e}}} } \right)\\
		%		&+ {\frac{{\phi A}}{{n(e - 1)}}\sum\limits_{e \in N_v} {\frac{{g_r^k(e)d_r}}{{c_e}}}},
		%		\end{aligned}\]
		\STATE
		${p_e} \leftarrow {p_e}\left( {1 + \frac{{{f^*_e}{d_r}}}{{{c_e}}}} \right){\rm{ + }}\frac{\phi }{{m}}\frac{{{f^*_e}{d_r}}}{{{c_e}}}$,
		\STATE where $\phi >0$.
		\ENDFOR
		\ENDIF
		\ENDWHILE
		\STATE \textbf{Output:} $X_r^f$
	\end{algorithmic}
\end{algorithm}

\subsection{Online Algorithm with Flow Shifting}
The modified online algorithm, as presented in \text{Algorithm~\ref{alg_online2}}, has a similar structure to \text{Algorithm \ref{alg_online}}.
The parameters $\eta$, $\sigma$ (and $\sigma:=\max{\{\eta,\mu\}}$) will be determined  in Section~\ref{sec:kernel:amcm}.
%Suppose that an approximate min-cost computation model 

\textbf{Theorem 4:}
\text{Algorithm \ref{alg_online2}} is an all-or-nothing, non-preemptive, monotone, and $\left\{ {O\left( 1 \right),O\left( {\log m} \right)} \right\}$-competitive,  more specifically $\left\{ {1 + \frac{\phi}{\sigma} ,\sigma\log \left( {\frac{{Bm}}{\phi } + 1} \right)} \right\}$-competitive, online algorithm.

\textit{Proof:}
See Appendix.
\qed

It is obvious that Algorithm \ref{alg_online2} has lower competitve ratio but higher violation ratio than Algorithm \ref{alg_online}.
However, the difference can only be reflected in theoretical analysis.

	\begin{algorithm}[!h]
	\caption{Online Algorithm for MISNC with Flow Shifting}\label{alg_online2}
	\begin{algorithmic}
		\STATE \textbf{Input:} $p_e \leftarrow 0,\forall e$
		\WHILE{Upon the arrival of request $r$}	 
		\STATE Invoke \textbf{Algorithm \ref{alg_amcm}} to compute the min-cost unit MCF from $s_r$ to $T_r$ in $G$ under the link price system $p$. Denote $f^*$ as the min-cost unit MCF. 
		\IF{$L>\lambda$}
		\STATE Reject $r$;
		\ELSE
		\STATE Accept $r$;	
		%		\STATE Let $k^* = \arg \mathop {\min }\limits_k \sum\limits_e {g_r^k(e)l(e)}$
		\STATE $X_r^{{f^*}} \leftarrow 1$ 
		\STATE ${z_r} \leftarrow {d_r}\left( {1 - \frac{L}{\sigma}} \right)$
		\FOR{$\forall e\in f^*$}
		%		\STATE \[\begin{aligned} 
		%		w_v \leftarrow w_v&\left( {1+ \sum\limits_{e \in N_v} {\frac{{g_r^k(e)d_r}}{{c_e}}} } \right)\\
		%		&+ {\frac{{\phi A}}{{n(e - 1)}}\sum\limits_{e \in N_v} {\frac{{g_r^k(e)d_r}}{{c_e}}}},
		%		\end{aligned}\]
		\STATE
		%		${p_e} \leftarrow {p_e}\left( {1 + {\rho _e}} \right) + \frac{\phi }{m}{\rho _e}$,
		${p_e} \leftarrow {p_e}\left( {1 + \frac{{{f^*_e}{d_r}}}{{\sigma{c_e}}}} \right){\rm{ + }}\frac{\phi }{{m}}\frac{{{f^*_e}{d_r}}}{{\sigma{c_e}}}$,
		\STATE where $\phi >0$.
		\ENDFOR
		\ENDIF
		\ENDWHILE
		\STATE \textbf{Output:} $X_r^f$
	\end{algorithmic}
\end{algorithm}

\section{Min-Cost SISNC Computation}\label{sec:kernel}
For the primal-dual algorithms in both offline and online cases, efficiently computing a min-cost MCF for a given price system is the key module in each iteration. 
The min-cost NC muilticast problem can be solved, either exactly or approximately, in many ways, such as optimization decomposition techniques or general LP solvers. 
In this section, we give an exact version of \textit{Min-Cost SISNC Computation} in Algorithm \ref{alg_mcm} and an approximate version in Algorithm \ref{alg_amcm}.
The former is integrated into the FPTAS and the online algorithm
while the latter is only for the online algorithm. 

\subsection{Exact Min-Cost SISNC Computation}\label{sec:kernel:mcm}
According to \cite{cl,mit}, the min-cost SISNC problem can be formulated as the following LP.
\begin{align}
\textbf{min}~&\sum\limits_{e}{{{p}_{e}}{{f}_{e}}} & (MNC) \nonumber\\
%(MNC)~\textbf{min}~&\sum\limits_{e}{{{p}_{e}}{{f}_{e}}} \nonumber\\
\textbf{s.t.}~
\label{eq:mnc:f} &\sum\limits_{p\in {{P}_{i}}}{f(p)}=1,&\forall i. \\
\label{eq:mnc:c} &{{f}_{i,e}}:=\sum\limits_{e\in p\in {{P}_{i}}}{f(p)}\le {{f}_{e}},&\forall i,\forall e. \\
\label{eq:mnc:capacity} &{f_e} \le \frac{{{c_e}}}{{{d_r}}},&\forall e. \\
&f(p),&\forall p. \nonumber\\
&{{f}_{e}},&\forall e. \nonumber
\end{align}

Constraint (\ref{eq:mnc:f}) follows the rule of flow conservation with unit flow request. 
The index $i$ indicates a unicast from $s_r$ to one of the multicast receivers $i \in {T_r}$
and $P_i$ represents the path set of this unicast.
Constraint (\ref{eq:mnc:c}) means that the \textit{conceptual link flow} $f_{i,e}$ is constrained by the \textit{actual link flow} $f_e$. This constraint characterizes the NC multicast.
Constraint (\ref{eq:mnc:capacity}) ensures that the actual link flow $f_e$ is constrained by the scaled link capacity.
Any feasible solution $f_e$ of problem (\textit{MNC}) constitutes a unit multicast flow polytope ${{\Pi }_{r}}$ for request $r$.
%Constraint (\ref{eq:p:nc}) is a relaxation of (\ref{eq:xf:max}) \cite{li:ton}.
%Note that $\theta  \ge 0$ and $X_{ij}^f \ge 0$ are automatically satisfied due to constraints (\ref{eq:p:c})(\ref{eq:p:nc})(\ref{eq:p:xftl}).
%$\theta > 1$ implies an \emph{overloaded} network.

\begin{algorithm}[!ht]
	\caption{Exact Min-Cost SISNC Computation}\label{alg_mcm}
	\begin{algorithmic}
		\STATE \textbf{Input:} $G$; $r$; $p$
		\STATE Solve the problem (\textit{MNC}) by invoking the decomposition algorithm in \cite{mit} or using an LP solver.
		\STATE The minimum cost of the unit MCF is:
		\STATE \quad $L := \mathop {\min }\limits_{f \in V({\Pi _r})} \sum\limits_e {{p_e}{f_e}}$
		\STATE The min-cost unit MCF is:
		\STATE \quad $f^* := \arg \mathop {\min }\limits_{f \in V({\Pi _r})} L$
		\STATE \textbf{Output:} $f^*$; $L$
	\end{algorithmic}
\end{algorithm}

\begin{algorithm}[!ht]
	\caption{Approximate Min-Cost SISNC Computation}\label{alg_amcm}
	\begin{algorithmic}
		\STATE \textbf{Input:} $G$; $r$; $p$
		\STATE Solve the problem (\textit{MNC}) by invoking the decomposition algorithm in \cite{mit} or using an LP solver.
		\STATE The minimum cost of the unit MCF is:
		\STATE \quad $L := \mathop {\min }\limits_{f \in V({\Pi _r})} \sum\limits_e {{p_e}{f_e}}$
		\STATE The min-cost unit MCF is:
		\STATE \quad $f^* := \arg \mathop {\min }\limits_{f \in V({\Pi _r})} L$
		
		\FOR{$\forall i\in {T_r}$}
		\STATE
		Shift the path flows smaller than $\frac{1}{{2{w_i}^2}}$ to other flow paths from $s_r$ to $i$ according to the rules in \cite{high_demands}.
		\ENDFOR
		
		\STATE Update $f^*$ and $L$.
		\STATE \textbf{Output:} $f^*$; $L$
	\end{algorithmic}
\end{algorithm}

\subsection{Approximate Min-Cost SISNC Computation}\label{sec:kernel:amcm}
\text{Algorithm \ref{alg_amcm}} is in essence an approximate algorithm. 
It produces a solution where the min-cost value $L$ is at most $\eta$ times the optimum
and the link loads of $f^*$ are at most $\mu$ times the link capacities.
In the following, we call it a \textit{$(\eta,\mu)$-criteria} algorithm.
Based on this criteria, \text{Algorithm \ref{alg_mcm}} can be regarded as a $(1,1)$-criteria algorithm.

The core of \text{Algorithm \ref{alg_amcm}} is a simple \textit{flow shifting} manipulation based on the outputs of \text{Algorithm \ref{alg_mcm}} .
Denote the number of paths from $s_r$ to $i \in {T_r}$ by $w_i$.
For each unicast from $s_r$ to $i$, shift the path flows smaller than $\frac{1}{{2{w_i}^2}}$ to other flow paths according to \cite{high_demands}.
Since different unicast flows in an NC multicast can overlap with each other when sharing the link capacity, implied by constraints (\ref{eq:mnc:c}) and (\ref{eq:mnc:capacity}),
the min-cost approximation ratio and the capacity violation ratio are identical to the unicast MCF case in \cite{high_demands}, i.e., $\eta=\mu=2$.

\textbf{Definition 4:}
Define the \textit{granularity} ${F_{{\rm{min}}}}$ as the smallest link load value of $f^*$ \cite{high_demands}.
In both of Algorithm \ref{alg_mcm} and Algorithm \ref{alg_amcm}, ${F_{{\rm{min}}}}$ is a constant.

\textbf{Theorem 5:}
In \text{Algorithm \ref{alg_amcm}}, ${F_{{\rm{min}}}}$ is lower bounded by $\frac{1}{{2{w_{\max }}^2}}$, i.e.:
${F_{{\rm{min}}}}{\rm{ \ge }}\frac{1}{{2{w_{\max }}^2}}$, where ${w_{\max }} = \mathop {\max }\limits_{i \in {T_r}} {w_i}$.
Further, \text{Algorithm \ref{alg_amcm}} is a $(2,2)$-criteria algorithm.

\textit{Proof:}
Refer to \cite{high_demands}.
\qed

%\textit{Remark:}
As can be seen in the Appendix, the introduction of \textit{granularity} is to upper bound the dual link price variables,
and implies that a flow between a node pair should not be split into too many tiny sub flows.
This generally holds in practice and is indispensable to prove the primal feasibility of the online algorithm. 
The flow shifting operation is designed to further lower bound the granularity for theoretical analysis 
and thereby makes Algorithm \ref{alg_online2} less practical due to large implementation and computation overheads \cite{high_demands}.
%In fact, \cite{high_demands} provides a much more complicated approximated Mincost Computation Module to exactly lower bound the granularity, where tiny flows are shifted to other paths for each unicast request.
%According to Theorem 1, all the unicast flows in a multicast with ISNC \textit{overlap} with each other when competing for link capacities. 
%Thus, the online algorithm in this paper can also be adapted in a similar manner by shifting tiny flows to other paths for each unicast in a multicast with ISNC.
%The results and proofs are rather similar to those in \cite{high_demands}.
%We omit the details here due to space limitation.

\section{Numerical Examples}
\begin{figure}[!h]
	\centering
	\includegraphics[angle=0, width=0.4\textwidth]{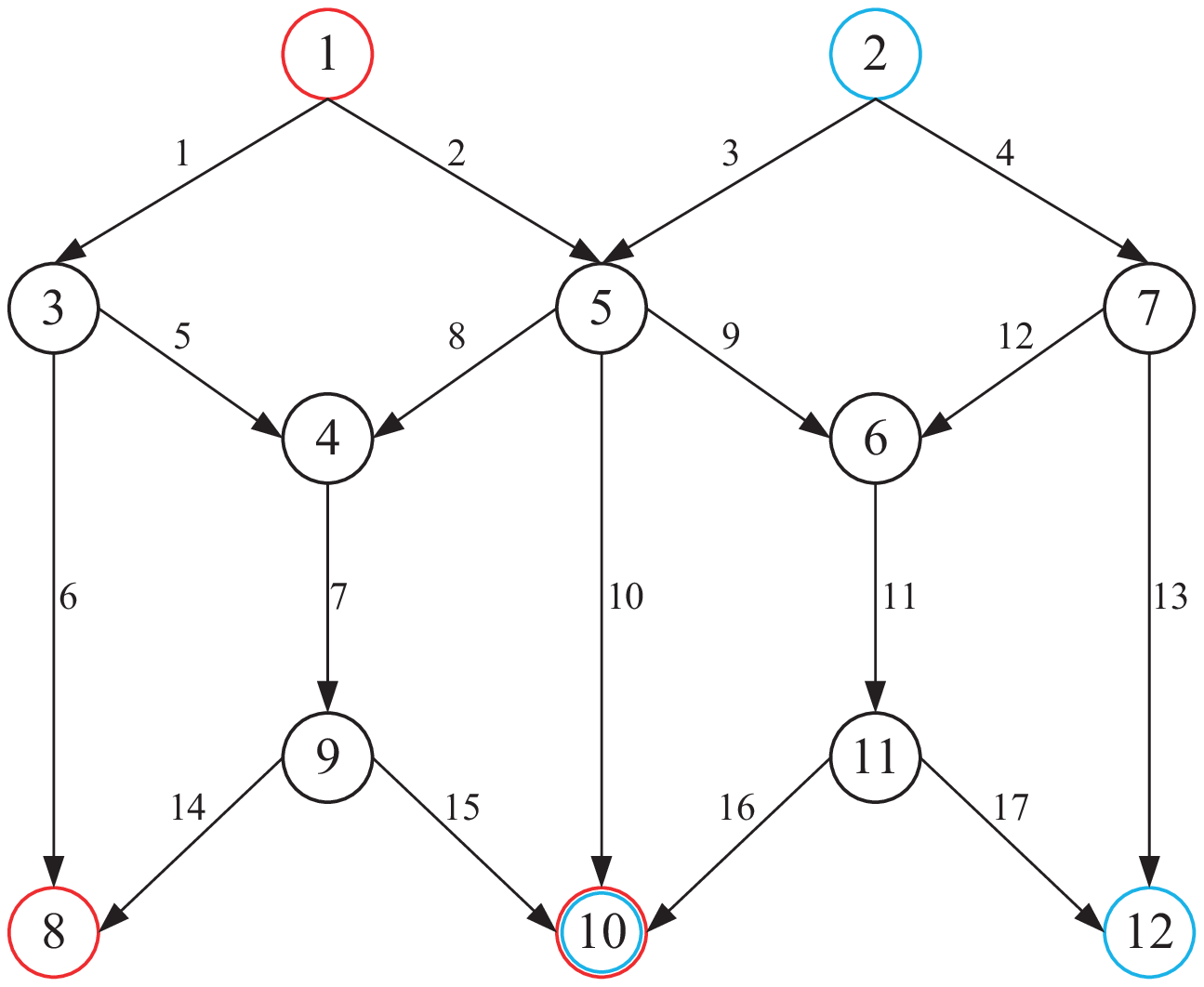}
	\caption{Testing topology \cite{cl}. Link indices are the numbers next to the links.}
	\label{fg:topo}
\end{figure}

Considering that the proposed algorithms prove to run in polynomial time and have provable worst case performance
and that we hope to clearly show the flow distribution over all the network links for ease of analysis,
we use a prototype network for the MISNC problem, see Fig.~\ref{fg:topo}, which is modified from the classical \textit{butterfly network} \cite{cl,comnet:multicast}. 
We use this network because it can clearly shows the nature of ISNC.
That is, it is enough to characterize the link sharing within the same multicast session and the link competition between different multicast sessions.
More importantly, it is also convenient to verify the correctness of our solutions since the optimum can be easily computed.

All network links have the same capacities 100 and the same weights 1.
There are two type of multicast requests $r_1$ and $r_2$ to be transmitted.
The source node sets and sink node sets are $s_{r_1}=\{1\}$, $T_{r_1}=\{8,10\}$ and $s_{r_2}=\{2\}$, $T_{r_2}=\{10,12\}$, respectively.

\subsection{Offline Setting}
\begin{figure}[!h]
	\centering
	\includegraphics[angle=0, width=0.43\textwidth]{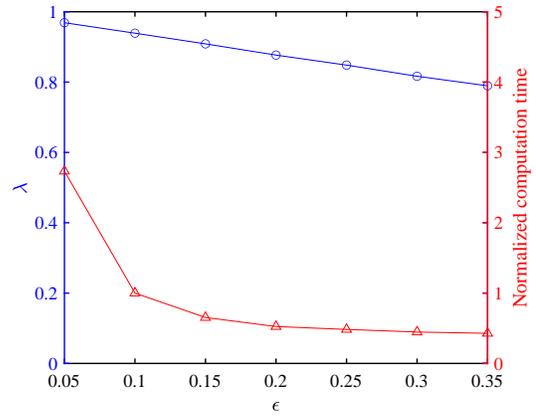}
	\caption{FPTAS. The optimal value of $\lambda$ is 1. The computation time under $\epsilon=0.1$ is used to normalize all the other settings of $\epsilon$.}
	\label{fg:fptas}
\end{figure}

Fig.~\ref{fg:fptas} validates the parameter $\epsilon$ in terms of the routing performance metric $\lambda$ and the computation cost metric.
The \textit{normalized computation time} is defined as the ratio of real computation time to the real computation time when $\epsilon=0.1$ (a commonly used setting in literature).
In the offline setting, the request size is set as $d_{r_1}=d_{r_2}=150$.
Thus, the optimal value of $\lambda$ is exactly 1.

In Fig.~\ref{fg:fptas}, we can see that when $\epsilon$ decreases, $\lambda$ grows linearly while the computation time grows exponentially.
The setting $\epsilon=0.1$ can reach a tradeoff between routing performance and computation cost.
\subsection{Online Setting}
\begin{figure}[!h]
	\centering
	\includegraphics[angle=0, width=0.43\textwidth]{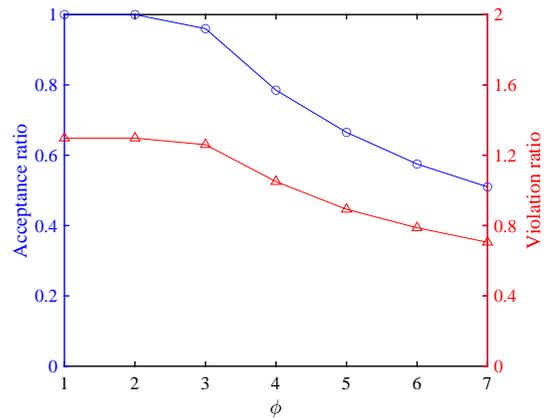}
	\caption{Online algorithm. A violation ratio larger than one means that the flow value on some link exceeds its capacity. }
	\label{fg:online}
\end{figure}

Fig.~\ref{fg:online} validates the parameter $\phi$ in terms of the routing performance metric \textit{acceptance ratio} and the resource cost metric \textit{violation ratio}.
The \textit{acceptance ratio} is defined as the ratio of accepted number of requests to the total request number.
The \textit{violation ratio} is defined as the maximum ratio of the real flow amount on a link to its capacity over all links.
It can be obviously seen that the performance bottleneck is link 10, which dominates the violation ratio.  

In the online setting, we generate 100 requests for request type 1 and 100 requests for request type 2.
Without loss of generality, we make these requests have equal size $d_{r_1}=d_{r_2}=1.5$. 
All the 200 requests are randomly ordered and will be injected to the network one by one.

We reuse this request sequence in different $\phi$ values for the sake of fairness.
Each multicast request is injected in the network at the ingress node 1 or 2, and is accepted if received by node set \{8,10\} or \{10,12\}, respectively.

From Fig.~\ref{fg:online}, if the capacity constraints are fully respected, the online algorithm when $\phi=4$ can reach an acceptance ratio of around 0.8, which is entirely satisfactory for an online setting.
\subsection{Comparison between the Two Settings}
The two algorithms share a common primal-dual structure. 
The requests are processed one by one over a long time scale.
We make delicate parameter settings for ease of comparison when testing the two algorithms.
Specifically, the total request size 100 in the online setting equals to that in the offline setting.

\begin{figure}[!h]
	\centering
	\includegraphics[angle=0, width=0.49\textwidth]{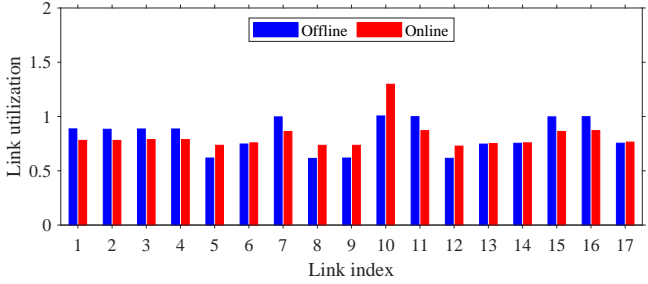}
	\caption{Comparison of link load. Offline ($\epsilon=0.1$). Online ($\phi=1$).}
	\label{fg:linkload}
\end{figure}

Fig.~\ref{fg:linkload} compares the link utilization between the two algorithms.
As a whole, the link utilization of the online algorithm is quite close to that of the FPTAS.
The online algorithm when $\phi=1$ accepts all the 200 requests at the cost of a noticable increase in link utilization, and thus in a violation ratio of around 1.2, with the bottleneck link 10.

%In Fig.~\ref{fg:example:comb}, we fix $L$ and study the effect of node combinations within $A$.
%There are $\binom{\left| L \right|}{\left| A \right|}$ combinations for inflow and $\binom{\left| N \right|-\left| L \right|}{\left| A \right|-\left| L \right|}$ combinations for outflow.
%Fig.~\ref{fg:wired:plus} shows that when there is no limitation on $d_i$ (or similarly, $d_i \gg u_i$), 
%a smaller ${\left| A \right|}$ leads to a strictly larger $r_{\rm max}$ than does a greater ${\left| A \right|}$, regardless of the node combinations. 
%In the other three figures where $d_i$ are limited, a smaller ${\left| A \right|}$ can only guarantee a possibly larger $r_{\rm max}$, not necessarily always so, because the node combination has a significant effect.

%\begin{figure*}[!ht]
%	\centering
%	\subfloat[Case I]{\includegraphics[width=0.45\textwidth]{mcf_nc_fptas}%
%		\label{fig:conv}}
%	\hfil
%	\subfloat[Case I]{\includegraphics[width=0.45\textwidth]{mcf_nc_online}%
%		\label{fig:conv:lp}}
%	\caption{Convergence validation. The dual decomposition is performed at: (a) Both level-1 and level-2. (b) Only level-1.}
%	\label{fig:converge}
%\end{figure*}

\section{Conclusion}
In this paper, we investigate the MISNC problem in terms of throughput maximization.
With the help of the extreme flow decomposition technique, we propose an FPTAS and online primal-dual algorithms for offline and online settings, respectively.
All the proposed algorithms have provable performance guarantees.
Numerical examples show that the FPTAS can asymptotically approach the optimum,
and the online algorithm can achieve a desirable trade-off between the routing performance and the resource constraints violation.

%This work can be extended in several directions.
%First, we can introduce two approximation parameters in the min-cost computation module, one for the capacity violation of links and the other for the object function value, just as \cite{high_demands} does. For clarity in the presentation, we omit these parameters in the proposed algorithms.

As one of our future works, we aim to develop a min-cost version of FPTAS to replace the SISNC computation module used in this paper.
We believe this can further accelerate the whole computation process.
%We believe this can accelerate the computation 

\section*{Appendix}
\subsection{Proofs for Algorithm~\ref{alg_fptas}}\label{sec:appendix:fptas}
\textbf{Lemma 1:} When the FPTAS terminates, the primal solution needs to be scaled by a factor of at most
${\log _{1 + \epsilon }}\frac{1}{\delta }$ to ensure primal feasibility (i.e., satisfying link capacity constraints).

\textit{Proof:}
Serialize all the iterations of all the phases into $l$ iterations.
Define the \textit{flow scaling factor} of link $e$ as:
\[\kappa : = \sum\limits_{i = 1}^l {\frac{{{d_r^{(i)}}f_e^{(i)}}}{{{c_e}}}} \]

According to the update rule of $p_e$, we have:
\[p_e^{\rho  - 1,K} = \frac{\delta }{{{c_e}}}\prod\limits_{i = 1}^l {\left( {1 + \epsilon \frac{{{d_r^{(i)}}f_e^{(i)}}}{{{c_e}}}} \right)} \]

Using the \textit{Taylor Formula}, the inequality ${(1 + x)^a} \le 1 + ax,\forall x,\forall a \in [0,1]$ holds.
Setting $x = \epsilon$ and $a = \frac{{{d_r^{(i)}}f_e^{(i)}}}{{{c_e}}} \le 1$, we have:
\begin{align*}
1 > {c_e}p_e^{\rho  - 1,K} &\ge \delta \prod\limits_{i = 1}^l {{{\left( {1 + \epsilon } \right)}^{\frac{{{d_r^{(i)}}f_e^{(i)}}}{{{c_e}}}}}}  \nonumber\\
&{ = }\delta {\left( {1 + \epsilon } \right)^{\sum\limits_{i = 1}^l {\frac{{{d_r^{(i)}}{f_e^{(i)}}}}{{{c_e}}}} }} \nonumber\\
&{ = }\delta {\left( {1 + \epsilon } \right)^\kappa } \nonumber
\end{align*}

Hence, the lemma is proven.
\[\kappa  < {\log _{1 + \epsilon }}\frac{1}{\delta }\]

\qed

\textbf{Lemma 2:}
At the end of $\rho$ phases in the FPTAS, we have
\[\frac{\beta }{{\rho  - 1}} \le \frac{\epsilon }{{(1 - \epsilon )\ln \frac{{1 - \epsilon }}{{m\delta }}}}\]

\textit{Proof:}
For a given $p$, $z_j$ is the minimum cost of shipping $d_j$ units of flow from $s_j$ to $T_j$ under price function $p$; henceforth denoted by ${\rm{mincost}}_j({p})$, i.e.:
\[{\rm{mincost}}_j({p}) = \mathop {\min }\limits_{f \in V({\Pi _j})} \sum\limits_e {p_e{f_e}} \]

Define
\[\alpha (p): = \sum\limits_{j = 1}^K {{d_j}{\rm{mincos}}{{\rm{t}}_j}(p)} \]
%\begin{small}
\begin{align*}
D({p^{i,j}})
& = \sum\limits_e {{c_e}p_e^{i,j}} \nonumber\\
& = D(p_e^{i,j - 1}) + \epsilon {d_j}\sum\limits_e {p_e^{i,j - 1}f_e^{i,j}} \nonumber\\
& = D({p^{i,j - 1}}) + \epsilon {d_j}{\rm{mincos}}{{\rm{t}}_j}({p^{i,j - 1}}) \nonumber
\end{align*}
%\end{small}

We now sum over all iterations during phase $i$ to obtain:
\begin{align*}
D({p^{i,K}})
& {\le } D({p^{i,0}}) + \epsilon \sum\limits_{j = 1}^K {{d_j}{\rm{mincos}}{{\rm{t}}_j}({p^{i,K}})} \nonumber\\
&{ = } D({p^{i,0}}) + \epsilon \alpha ({p^{i,K}}) \nonumber\\
&{ = } D({p^{i - 1,K}}) + \epsilon \alpha ({p^{i,K}}) \nonumber
\end{align*}

Since $\beta  \le \frac{{D({p^{i,K}})}}{{\alpha ({p^{i,K}})}}$, we have:
\[D({p^{i,K}}) \le \frac{{D({p^{i - 1,K}})}}{{1 - \frac{\epsilon }{\beta }}}\]

Using the initial value $D({p^{1,0}}) = m\delta $, we have for $i \ge 1$
\[D({p^{i,K}}) \le \frac{{m\delta }}{{1 - \epsilon }}{e^{\frac{{\epsilon (i - 1)}}{{\beta (1 - \epsilon )}}}}\]

The last step uses the assumption that $\beta \ge 1$. 
The procedure stops at the first phase $\rho$ for which
\[1 \le D({p^{\rho ,K}}) \le \frac{{m\delta }}{{1 - \epsilon }}{e^{\frac{{\epsilon (\rho  - 1)}}{{\beta (1 - \epsilon )}}}}\]
which implies that
\[\frac{\beta }{{\rho  - 1}} \le \frac{\epsilon }{{(1 - \epsilon )\ln \frac{{1 - \epsilon }}{{m\delta }}}}\]
\qed

\textbf{Proof of Theorem 2:} 
The analysis of the algorithm proceeds as in \cite{fptas2007,p2p}. 
Since the constraints are based on extreme flows and are related to MISNC, 
the analysis is different.% in two aspects: 1) the node weights include the capacity information. 2) existence of ECMP and Loop.

\textit{1) Approximation ratio:}
We use $\gamma$, the ratio of the dual to the primal objective function values, to upper bound and represent the approximation ratio. 
Then we have
\[\frac{\mathcal{D}}{\mathcal{P}}:=\gamma <\frac{\beta }{\rho -1}{{\log }_{1+\epsilon }}\frac{1}{\delta }\]

Substituting the bound on $\frac{\beta }{\rho -1}$ from Lemma 2, we have
\[\gamma <\frac{\epsilon {{\log }_{1+\epsilon }}\frac{1}{\delta }}{(1-\epsilon )\ln \frac{1-\epsilon }{m\delta }}=\frac{\epsilon }{(1-\epsilon )\ln (1-\epsilon )}\frac{\ln \frac{1}{\delta }}{\ln \frac{1-\epsilon }{m\delta }}\]

Setting $\delta ={{\left( \frac{1-\epsilon }{m} \right)}^{\frac{1}{\epsilon }}}$, we have $\gamma \le {{(1-\epsilon )}^{-3}}$.

Equating the desired approximation factor $(1+\omega )$ to this ratio and solving for $\epsilon$, we get the value of $\epsilon$
stated in the theorem.

\textit{2) Running time:}
Using weak-duality from linear programming theory, we have
\[1\le \gamma <\frac{\beta }{\rho -1}{{\log }_{1+\epsilon }}\frac{1}{\delta }\]

Then the number of phases $\rho$ is upper bounded by
\[\rho =\left\lceil \frac{\beta }{\epsilon }{{\log }_{1+\epsilon }}\frac{m}{1-\epsilon } \right\rceil \]

Note that the number of phases derived above is under the assumption $\beta  \le 2$.
The case $\beta > 2$ can be recast as $1 \le \beta  \le 2$
by scaling the link capacities and/or request sizes using the same technique in Section 5.2 of \cite{fptas2007}. 
Then, the number of phases is at most $2\rho\log |R|$.
We omit the details here.

Since each phase contains $|R|$ iterations, the total number of steps is at most
%\[(2K\log K+m)\left\lceil \frac{\beta }{\epsilon }{{\log }_{1+\epsilon }}\frac{m}{1-\epsilon } \right\rceil \]
%\[{2|R|\log |R|\left\lceil \frac{\beta }{\epsilon }{{\log }_{1+\epsilon }}\frac{m}{1-\epsilon } \right\rceil} \equiv {O\left( \frac{|R|\log |R|}{\epsilon }{{\log }_{1+\epsilon }}\frac{m}{1-\epsilon } \right)}\]
\[{2|R|\log |R|\rho} \equiv {O\left( \frac{|R|\log |R|}{\epsilon }{{\log }_{1+\epsilon }}\frac{m}{1-\epsilon } \right)}\]

Multiplying the above by $T_{\rm{NC}}$, the running time of each iteration, proves the theorem.

\qed

\subsection{Proofs for Algorithm~\ref{alg_online}}\label{sec:appendix:online}

\textbf{Proof of Theorem 3:} 
The online algorithm is by nature an approximation algorithm, and the performance guarantee can be proved in three steps as in \cite{sr}.
Since we use a multipath meta-structure (MCF) rather than a single-path meta-structure (path or walk) in the innermost loop of the algorithm, the proof, especially the competitive ratio analysis, is somewhat different.

\textit{1) Dual feasibility:} 
We first show that the dual variables $p_e$ and $z_r$ generated in each iteration by the algorithm are feasible.

Let $f^*$ denote the intermediate node that minimizes
${\sum\limits_e {{p_e}{f_e}} }$.

Setting 
${z_r} = {d_r}\left( {1 - L } \right)$
makes the constraints
${z_r} \ge {d_r}\left( {1 - \sum\limits_e {{p_e}{f_e}} } \right)$
hold for all feasible extreme flows.
The subsequent increase in $p_e$ will always maintain the inequality since $z_r$ does not change.

\textit{2) Competitive ratio:}
During the iteration where request $r$ is accepted, the increase in the primal function is:
\[\Delta \mathcal{P}=d_r\]
and the increase in the dual function is:
\[\Delta \mathcal{D}=z_r{+}\sum\limits_e {c_e}{\Delta p_e}. \]

Combining
$z_r = {d_r}\left( {1 - L} \right)$
and
\begin{align*}
\sum\limits_e {{c_e}\Delta {p_e}}  &= \sum\limits_e {{c_e}\left( {{p_e}\frac{{f_e^*{d_r}}}{{{c_e}}} + \frac{\phi }{m}\frac{{f_e^*{d_r}}}{{{c_e}}}} \right)} \nonumber\\
& = {d_r}\sum\limits_e {\left( {{p_e}f_e^* + \frac{\phi }{m}f_e^*} \right)}, \nonumber
\end{align*}
we have
\begin{align*}
{z_r}{\rm{ + }}\sum\limits_e {{c_e}\Delta {p_e}}  &= {d_r}\left( {1 + \frac{{\phi \sum\nolimits_e {f_e^*} }}{m}} \right) \nonumber\\
& \le {d_r}(1 + \phi ) \nonumber
\end{align*}

The last inequality holds since $\sum\limits_e {f_e^*}  \le m$.

Therefore, the competitive ratio can be calculated as:
%\[\frac{{\Delta P}}{{\Delta D}}{\rm{ = }}\frac{{d_r}}{{z_r{\rm{ + }}\sum\nolimits_e  {c_e}{\Delta p_e} }} \ge \frac{{1}}{{1 + \phi }} \equiv O(1)\]
\[\frac{{\Delta \mathcal{D}}}{{\Delta \mathcal{P}}}{\rm{ = }}\frac{{z_r{\rm{ + }}\sum\nolimits_e  {c_e}{\Delta p_e} }}{{d_r}} \le {1 + \phi } \equiv O(1)\]

\textit{3) Primal feasibility:}  
We now show that the solution is \textit{almost}\footnote{Here, \textit{almost} means the link capacity constraint may be \textit{slightly}, more exactly \textit{logarithmically}, violated.}
primal feasible.
The key of this step is to give the lower bound and upper bound of the link price $p_e$.

Denote the link price $p_e$ after request $r$ has been accepted and processed by $p(e,r)$,
and the utilization of link $e$ as 
\[\rho (e,r): = \frac{{F(e,r)}}{{{c_e}}}\]

First, we prove an lower bound of $p(e,r)$:
\[p(e,r) \ge \frac{\phi }{m}\left[ {{e^{\rho (e,r)}} - 1} \right]\]

We use the induction method. 
According to the update rule of $p_e$, we have:
% 1-column
%\begin{small}
	\begin{align*}
	p(e,r) 
	&\ge p(e,r - 1)\left( {1 + \frac{{{d_r}f_e^*}}{{{c_e}}}} \right){\rm{ + }}\frac{\phi }{m}\frac{{{d_r}f_e^*}}{{{c_e}}}\nonumber\\
	& \ge \frac{\phi }{m}\left[ {{e^{\rho (e,r - 1)}} - 1} \right]\left( {1 + \frac{{{d_r}f_e^*}}{{{c_e}}}} \right){\rm{ + }}\frac{\phi }{m}\frac{{{d_r}f_e^*}}{{{c_e}}} \nonumber\\
	& = \frac{\phi }{m}\left[ {{e^{\rho (e,r - 1)}}\left( {1 + \frac{{{d_r}f_e^*}}{{{c_e}}}} \right) - 1} \right] \nonumber\\
	& \ge \frac{\phi }{m}\left[ {{e^{\rho (e,r)}} - 1} \right] \nonumber
	\end{align*}	
%\end{small}

The last inequality follows from:
\[\left( {1 + \frac{{{d_r}f_e^*}}{{{c_e}}}} \right) \approx {e^{\frac{{{d_r}f_e^*}}{{{c_e}}}}}\]
and
\[\rho (e,r) = \rho (e,r - 1) + \frac{{{d_r}f_e^*}}{{{c_e}}}\]

Second, we prove an upper bound of $p(e,r)$:
\[p(e,r) \le B\]

After request $r$ is accepted, the min-cost value $L \le 1$.
Then:
\[p(e,r - 1){F_{{\rm{min}}}} \le L \le 1\]

According to the update rule of $p(e,r)$ and $\frac{{{d_r}f_e^*}}{{{c_e}}} \le 1$, we have:
\[p(e,r) \le \frac{1}{{{F_{{\rm{min}}}}}} \cdot {\rm{2 + }}\frac{\phi }{m} \cdot 1 = \frac{1}{{{F_{{\rm{min}}}}}}{\rm{ + }}\frac{\phi }{m} \le \frac{{1 + \phi }}{{{F_{{\rm{min}}}}}}: = B\]

The last inequality follows from $m \ge 1 \ge {F_{{\rm{min}}}}$.
%Actually, the granularity cannot be infinitesimal.
%That is, it can be viewed as a bounded value however small it is. 
%Since the min-cost value $L$ is also a bounded value, the link price must be upper bounded by some constant $B$, i.e.:
%Let
%\[\frac{\phi }{m}\left( {{e^{\rho (e,r)}} - 1} \right) \le B\]
Combining the lower bound and the upper bound, we have:
\[\frac{F(e,r)}{c_e}:=\rho (e,r) \le \log \left( {\frac{{Bm}}{\phi } + 1} \right) \equiv O(\log m)\]
%We assume that the flow value of every link is larger than some constant, which is regarding the concept of \textit{granularity} \cite{high_demands}.
%We omit the details here due to space limitation. 
%The implication of this assumption is that a flow between a node pair cannot be split into too many small sub flows. 
%This assumption generally holds in practice and is indispensable to prove the primal feasibility of the online algorithm. 

%Intuitively, 

\qed

\subsection{Proofs for Algorithm~\ref{alg_online2}}\label{sec:appendix:online2}
\textbf{Proof of Theorem 4:} 
Althought the proof is similar to that of Theorem 2, we still present it here for completeness.

\textit{1) Dual feasibility:} 
Setting 
$z_r = {d_r}\left( {1 - \frac{L}{\sigma}} \right)$
makes the dual constraints
${z_r} \ge {d_r}\left( {1 - \sum\limits_e {{p_e}{f_e}} } \right)$
hold for all feasible extreme flows.
The subsequent increase in $p_e$ will always maintain the inequality since $z_r$ does not change.

\textit{2) Competitive ratio:}
During the iteration where request $r$ is accepted, the increases in the primal and dual objectives are:
\[\Delta \mathcal{P}=d_r\]
and
\[\Delta \mathcal{D}=z_r{+}\sum\limits_e {c_e}{\Delta p_e}. \]

Combining
$z_r = {d_r}\left( {1 - \frac{L}{\sigma}} \right)$
and
\begin{align*}
\sum\limits_e {{c_e}\Delta {p_e}}  &= \sum\limits_e {{c_e}\left( {{p_e}\frac{{f_e^*{d_r}}}{{\sigma{c_e}}} + \frac{\phi }{m}\frac{{f_e^*{d_r}}}{{\sigma{c_e}}}} \right)} \nonumber\\
& ={d_r}\sum\limits_e {\left( {\frac{{{p_e}f_e^*}}{\sigma} + \frac{\phi }{{m}}\frac{{f_e^*}}{\sigma}} \right)}, \nonumber
\end{align*}
we have
\begin{align*}
{z_r}{\rm{ + }}\sum\limits_e {{c_e}\Delta {p_e}}  &= {d_r}\left( {1 + \frac{{\phi \sum\nolimits_e {f_e^*} }}{\sigma m}} \right) \nonumber\\
& \le {d_r}(1 + \frac{\phi}{\sigma}) \nonumber
\end{align*}

Therefore, the competitive ratio can be calculated as:
%\[\frac{{\Delta P}}{{\Delta D}}{\rm{ = }}\frac{{{d_r}}}{{{z_r}{\rm{ + }}\sum\nolimits_e {{c_e}\Delta {p_e}} }} \ge \frac{1}{{1 + \phi /\sigma}} \equiv O(1)\]
\[\frac{{\Delta \mathcal{D}}}{{\Delta \mathcal{P}}}{\rm{ = }}\frac{{z_r{\rm{ + }}\sum\nolimits_e  {c_e}{\Delta p_e} }}{{d_r}} \le {1 + \frac{\phi}{\sigma} } \equiv O(1)\]

\textit{3) Primal feasibility:}  
Denote the \textit{scaled} utilization of link $e$ as 
\[\rho (e,r): = \frac{{F(e,r)}}{{\sigma{c_e}}}\]

First, we prove an lower bound of $p(e,r)$:
\[p(e,r) \ge \frac{\phi }{m}\left[ {{e^{\rho (e,r)}} - 1} \right]\]

We use the induction method. 
According to the update rule of $p_e$, we have:

% 1-column
%\begin{small}
\begin{align*}
p(e,r) 
&\ge p(e,r - 1)\left( {1 + \frac{{{d_r}f_e^*}}{{\sigma{c_e}}}} \right){\rm{ + }}\frac{\phi }{m}\frac{{{d_r}f_e^*}}{{\sigma{c_e}}}\nonumber\\
& \ge \frac{\phi }{m}\left[ {{e^{\rho (e,r - 1)}} - 1} \right]\left( {1 + \frac{{{d_r}f_e^*}}{{\sigma{c_e}}}} \right){\rm{ + }}\frac{\phi }{m}\frac{{{d_r}f_e^*}}{{\sigma{c_e}}} \nonumber\\
& = \frac{\phi }{m}\left[ {{e^{\rho (e,r - 1)}}\left( {1 + \frac{{{d_r}f_e^*}}{{\sigma{c_e}}}} \right) - 1} \right] \nonumber\\
& \ge \frac{\phi }{m}\left[ {{e^{\rho (e,r)}} - 1} \right] \nonumber
\end{align*}	
%\end{small}

The last inequality follows from:
\[\left( {1 + \frac{{{d_r}f_e^*}}{{\sigma{c_e}}}} \right) \approx {e^{\frac{{{d_r}f_e^*}}{{\sigma{c_e}}}}}\]
and
\[\rho (e,r) = \rho (e,r - 1) + \frac{{{d_r}f_e^*}}{{\sigma{c_e}}}\]

Second, we prove an upper bound of $p(e,r)$:
\[p(e,r) \le B\]

After request $r$ is accepted, $L \le \lambda$.
Then:
\[p(e,r - 1){F_{{\rm{min}}}} \le L \le \lambda\]

Following the update rule of $p(e,r)$ and $\frac{{{d_r}f_e^*}}{{\sigma{c_e}}} \le 1$, we have:
\[p(e,r) \le \frac{\lambda }{{{F_{{\rm{min}}}}}} \cdot {\rm{2 + }}\frac{\phi }{m} \cdot 1 = \frac{{2\lambda }}{{{F_{{\rm{min}}}}}}{\rm{ + }}\frac{\phi }{m} \le \frac{{2\lambda  + \phi }}{{{F_{{\rm{min}}}}}}: = B\]

%Actually, the granularity cannot be infinitesimal.
%That is, it can be viewed as a bounded value however small it is. 
%Since the min-cost value $L$ is also a bounded value, the link price must be upper bounded by some constant $B$, i.e.:
%Let
%\[\frac{\phi }{m}\left( {{e^{\rho (e,r)}} - 1} \right) \le B\]

Combining the lower bound and the upper bound, we have:
\[\rho (e,r) \le \log \left( {\frac{{Bm}}{\phi } + 1} \right) \equiv O(\log m)\]

Thus,
\[\frac{F(e,r)}{c_e} = \sigma\rho (e,r) \equiv O(\log m)\]

%We assume that the flow value of every link is larger than some constant, which is regarding the concept of \textit{granularity} \cite{high_demands}.
%We omit the details here due to space limitation. 
%The implication of this assumption is that a flow between a node pair cannot be split into too many small sub flows. 
%This assumption generally holds in practice and is indispensable to prove the primal feasibility of the online algorithm. 

%Intuitively, 

\qed

\end{document}